\documentclass[12pt,a4paper]{article}
\usepackage{mathtext}        
\usepackage{amssymb,amsfonts}
\usepackage{graphicx}
\usepackage{epsfig}
\usepackage{epsf}
\usepackage{rotate}
\parskip=\baselineskip

\def\beq#1{\begin{equation}\label{#1}}
\def\eeq{\end{equation}}
\def\beqa#1{\begin{eqnarray}\label{#1}}
\def\eeqa{\end{eqnarray}}

\def\myfrac#1#2{\left(\frac{#1}{#2}\right)}
\def\comment#1{\relax}
\makeatletter
\def\eqalign#1{\null\,\vcenter{\openup\jot\m@th
  \ialign{\strut\hfil$\displaystyle{##}$&$\displaystyle{{}##}$\hfil
      \crcr#1\crcr}}\,}
\def\eqalignleft#1{\null\,\vcenter{\openup\jot\m@th
  \ialign{\strut$\displaystyle{##}$\hfil&$\displaystyle{{}##}$\hfil
      \crcr#1\crcr}}\,}
\makeatother

\begin{document}

\begin{center}
{\large\bf\uppercase{X-ray luminosity function of 
low-mass X-ray binaries in galaxies}}

{\bf
K.A.Postnov(1,2), A.G.Kuranov(1)}
\end{center}

{\it

(1) Sternberg Astronomical Institute, Moscow, Russia

(2) University of Oulu, Finland
}

\begin{abstract}

X-ray luminosity function derived from observations of
X-ray sources in galactic bulges can be explained by prinicipal
evolutionary relations for mass accretion rate onto 
the compact object. The observed mean distribution of 
individual X-ray luminosities of galactic LMXB is satisfactorily
described by a symmetric 
quasi-Lorentzian curve. The flux variance for bright sources  
is found to be proportional to 
the mean luminosity. Such a distribution does not change the
slope of the power law luminosity function 
of the source population, which is expected from the 
dependence of the mass transfer rate on the mass of the Roche-lobe
filling non-degenerate optical component.  

\end{abstract}

\section{Introduction}

X-ray flux measurements from numeorus  
point-like sources in other galaxies
have recently become possible by Chandra and XMM-Newton
X-ray telescopes. Most of these sources 
are accreting compact stars in binaries. These observations
have important implications for population studies 
of X-ray binaries in galaxies. Grimm et al. (2993) and Gilfanov (2004) 
carried out a detailed statistical analysis of the observed 
X-ray luminosity function of X-ray sources observed by Chandra
and XMM. They constructed X-ray luminosity function of 
X-ray binaries in individual galaxies, derived the mean luminosity 
function and discussed its astrophysical implications. In our previous
paper (Postnov 2003) we noted that the universal power-law 
shape of X-ray luminosity function (XLF) of high-mass X-ray binaries  
obtained by these authors 
$dN/dL_x\sim L^{-1.6}$ in a wide range of X-ray liminosities 
$10^{34}-10^{40}$ erg/s can be readily explained by the universal
properties of accretion from stellar wind from the early-type 
component. Important points were the possibility to use 
the power-law dependence for mass-luminosity and mass-radius 
relations for early-type stars, as well as a natural assumption 
on the power-law initial stellar mass function (Salpeter law or
Miller-Scalo law). We also assumed that the mass accretion rate onto
the comact star is the principal parameter determining the X-ray luminosity:
$L_x\propto \dot M_a$. 

New observational data were used by Gilfanov (2004) (see also Kim and 
Fabbiano 2004) to construct XLF of low-mass X-ray binaries (LMXB). 
It turned out that the mean XLF for these sources, too, can 
be approximated by a univeersal function, which is, however, not
a single power-law, as in the case of HMXB. 

It is well known that X-ray binaries in general are variable. 
Moreover, many of them are transient sources, with X-ray luminosity changing 
by several times and even orders of magnitudes (e.g., X-ray novae) 
over different time scales. Typical X-ray observations do not last for more
than several tens kiloseconds, which in most cases less than the characteristic
time of significant luminosity changes in individual sources. So the question 
arises as to what exactly the observed mean XLF reflects, 
the instantaneous luminosity distribution of a collection of 
sources or the 
luminosity function of individual sources. 

In the present paper we first show that the observed dispersion of 
individual X-ray luminosities of galactic LMXB as obtained from the 
analysis of the RXTE ASM data can be ona average described by a 
symmetric Lorentzian distribution with the variance proortional to the 
mean flux. Nest we prove that 
such a distribution of accretion rate  $\dot M_a$ 
onto the compact 
object does not change a power-law character of XLF 
which is theoretically
calculated from the analysis of mass exchange rate 
$\dot M_o$ in close binaries. We arrive at the conclusion that 
the mean XLF of LMXB can be explained by evolutionary features 
of the mass transfer process in low-mass close binary systems:
the mass transfer rate due to binary's angular momentum removal
by gravitational waves shapes the low-luminosity end of of the
observed XLF $\propto L_x^{-1}$ ($L_x<2\times 10^{37}$ erg/s), 
while mass exchange rates driven by 
magnetic stellar wind from the late-type optical component can
reproduce the high-luminosity end of the observed XLF of LMXB
$\propto L_x^{-2}$ ($2\times 10^{37}<L_x<5\times 10^{38}$ erg/s).

\section{Power-law luminosity functions}

First of all, let us show on general grounds how power-law 
XLF in accreting binaries can be obtained. The basic assumption 
is the possibility to express the X-ray luminosity of a source
through the mass exchange rate $\dot M_o$ which is a powe-law
function of the optical star mass $M_o$ (Postnov 2003): 
\beq{1}
L_x \sim \dot M_a \sim \dot M_o \sim M_o^\alpha
\eeq
Below we shall consider the luminosity function per logarithmic 
interval $dN/d \ln L_x$. Under assumption (1) in 
the stationary situation we have: 
\beq{2}
\frac{dN}{d\ln L_x}=\frac{dN}{d\ln M_o}\frac{d\ln M_o}{d\ln L_x}
\propto L_x^{-\frac{\beta_{st}-1}{\alpha}}
\eeq
Here the stationary mass function for optical components was
utilized:
\beq{3}
 f_{st}(M)\equiv \frac{dN}{d M_o dt}\sim M_o^{-\beta_{st}}
\eeq
(for example, $\beta_{st}=2.35$ for the Salpeter mass function).
Consequently, a power-law shape for XLF is recovered: 
$$
\frac{dN}{d\ln L_x}\sim L_x^{-\Gamma}
$$
where the power-law index is 
$$
\Gamma=\frac{\beta_{st}-1}{\alpha}
$$
and depends only on the slope of the mass function and relationship (1). 

\subsection{Stationary mass function of optical components}

Two cases should be distinguished: The case of young massive close binary systems
in which the mass of the optical component virtually did not change 
from the original value, and teh case of old low-mass close binaries, 
where the mass of the optical star significantly diminished in 
the course of mass transfer. In the former case the mass fucntion of
the optical component does not alter significantly, so we can 
adopt its initial form:
\beq{5}
f(M)_{st}= f_o(M) \propto M^{-\beta}\,\qquad \beta_{st}=\beta
\eeq{}
In the latter case the stationary optical star mass distribution 
can be derived from the one-dimensional kinetic equation 
with stationary source $f_o(M)$
\beq{}
\frac{\partial }{\partial M}\left[f(M) \dot M \right] \propto f_o(M)
\eeq 
In our case $\dot M<0$ and we obtain
\beq{f_st}
\dot f_{st}(M)=\frac{\int_M^{M_{max}}\dot f_0(M')dM'}{\dot M}
%\,, \quad M>M_{min}
\eeq
where $M_{max}$ is the upper mass limit for stars. For power-law mass
fucntions its exact value is of minor importance. 

Substituting the power-law mass function  $f_o(M)\sim M^{-\beta}$ 
into Eq. (\ref{f_st}), we find
\beq{}
f_{st}(M) \propto M^{-\beta+1-\alpha}\,,
\eeq
that is the power-law index in the stationary stellar 
mass distribution becomes dependent on the mass exchange rate:
\beq{beta_st}
\beta_{st}=\beta-1+\alpha
\eeq
For example, for HMXB accreting from steallr wind 
$\alpha \approx 2.5$ (Postnov 2003) and for $\beta=2.35...2.5$ 
we get $\Gamma=0.54...0.6$
(the observed value is $\sim 0.6$).

\section{Effect of the luminosity function of individual
sources on the luminosity function of the source population}

\begin{table}%[h]
\caption{Low-mass X-ray binaries with short orbital periods.
For each source given are the orbital period 
$P(h)$ (in hours),
the mean ASM flux $F_x$ (in ASM count rate), 
the rms variance of the flux $\sigma$ (in ASM count rate), 
minimal ($F_{min}$) and maximal ($F_{max}$) ASM fluxes over
the entire time interval used in the analysis}
\begin{center}
\begin{tabular}{|r|l|c|c|c|c|c|}
\hline
N\% & Source    & $P(h)$ &$F_x$ & $\sigma$& $F_{min}$& $F_{max}$  \\
\hline
\hline
 1 &V1333 Aql    &  18.97&   2.31&   7.85& -13.85&  49.68\\
\hline
 2 &V821 Ara     &  14.86&   6.80&  13.51& -13.92&  72.57\\
\hline
 3 &V1408 Aql    &   9.33&   1.89&   1.23& -17.07&   6.95\\
\hline
 4 &LU TrA       &   9.14&   1.01&   1.06&  -4.59&   8.63\\
\hline
 5 &V691 CrA     &   5.57&   1.18&   1.33&  -8.62&  16.73\\
\hline
 6 &V926 Sco     &   4.65&  12.89&   3.38&  -9.70&  29.64\\
\hline
 7 &V2216 Oph    &   4.20&  17.34&   2.51&   0.00&  35.78\\
\hline
 8 &GR Mus       &   3.93&   1.93&   1.04&  -6.02&  23.36\\
\hline
 9 &V801 Ara     &   3.80&  10.90&   4.88&  -5.63&  34.92\\
\hline
10 &1705-4402    &   1.31&  11.67&   8.85&  -3.89&  35.94\\

\hline \hline
\end{tabular}
\end{center}
\end{table}

Before considering a more complicated case of low-mass close
binary systems, let us discuss the influence of individual source XLF 
on the XLF of the entire population. Let the accretion rate onto the compact object and hence X-ray 
luminosity $L_x$ of a source changes according to some
distribution function $F(x)$ (i.e., $F(x)$ gives the
probability to detect the X-ray luminosity in the range 
$[F(x),F(x+\Delta x)]$). Here the mean mass exchange rate
$\dot M_o$ is allowed to change or remain constant. Let  
$L_0$ be the X-ray luminosity corresponding to the mean accretion
rate. We assume that this rate represents the mass exchange
rate between the components determined by the system's evolutionary
state. It is easy to see that the observed XLF is the convolution
\beq{svertka}
dN/d L_x=
\frac{\int_{L_{min}}^{L_{max}} (dN/d L_0) F(L_x-L_0) dL_0}
{\int_{L_{min}}^{L_{max}} F(L_x-L_0) dL_0}
\eeq
We restrict ourselves by considering power-law XLFs, i.e. 
$dN/d\ln L_0\sim L_0^{-\Gamma}$. If the function $F(x-y)$
can be factorized in the form
\beq{F}
F(x-y)=F_1(y)F_2(x/y-1)\,,
\eeq
then the power-law XLF of the source population 
$dN/d\ln x\sim x^{-\Gamma}$ preserves when 
$F_1(y)$ is also a power-law function 
for arbitrary function 
$F_2$ (all necessary integration conditions are assumed). 
The proof readily follows from changing variables 
$t=x/y-1$ in integrals
(\ref{svertka}) and setting $L_{min}\to 0$, $L_{max}\to\infty$.

\begin{figure}
\begin{center}
\includegraphics[angle=270, totalheight=1.4in]{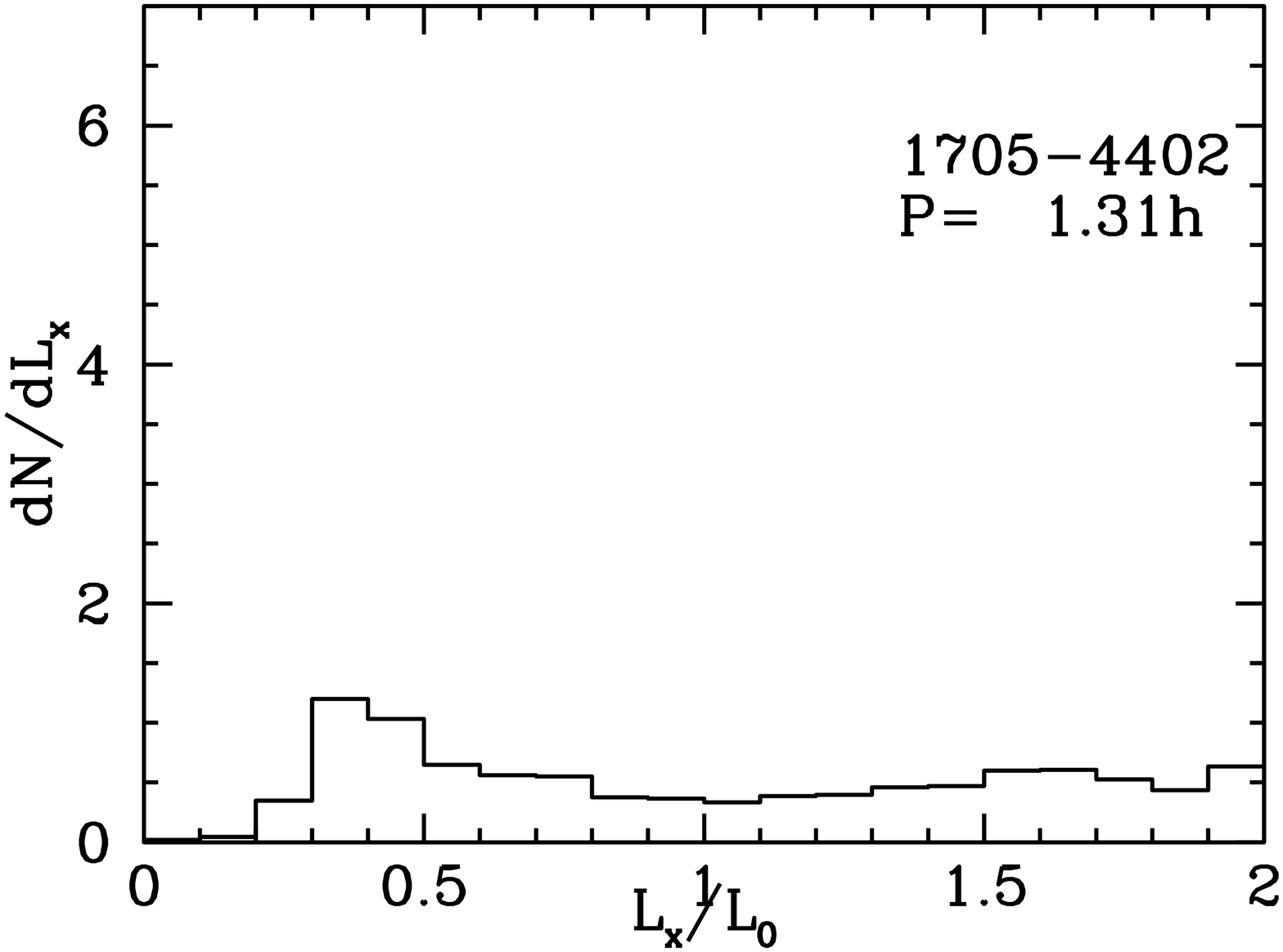}
\includegraphics[angle=270, totalheight=1.4in]{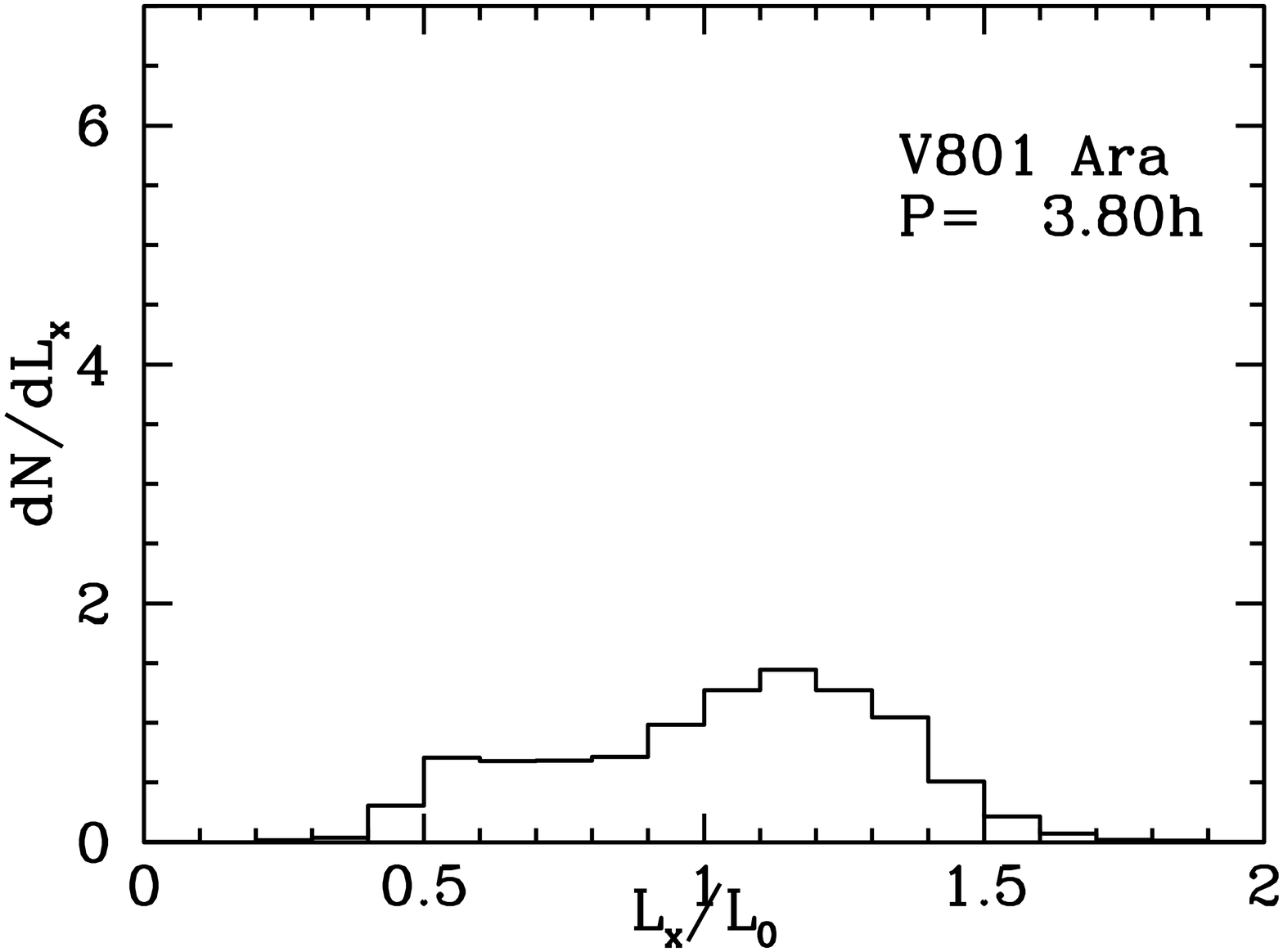}
\includegraphics[angle=270, totalheight=1.4in]{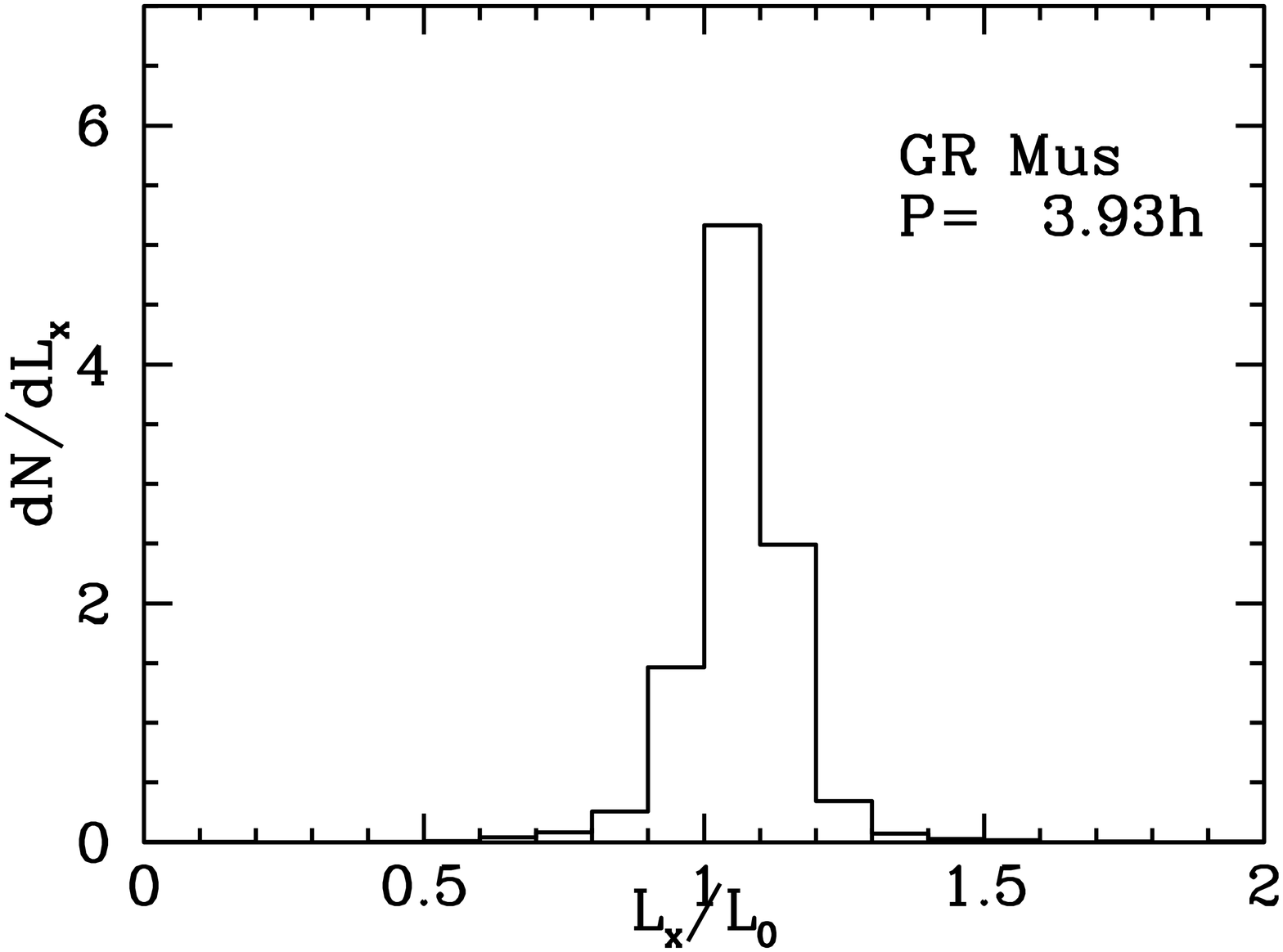}
\includegraphics[angle=270, totalheight=1.4in]{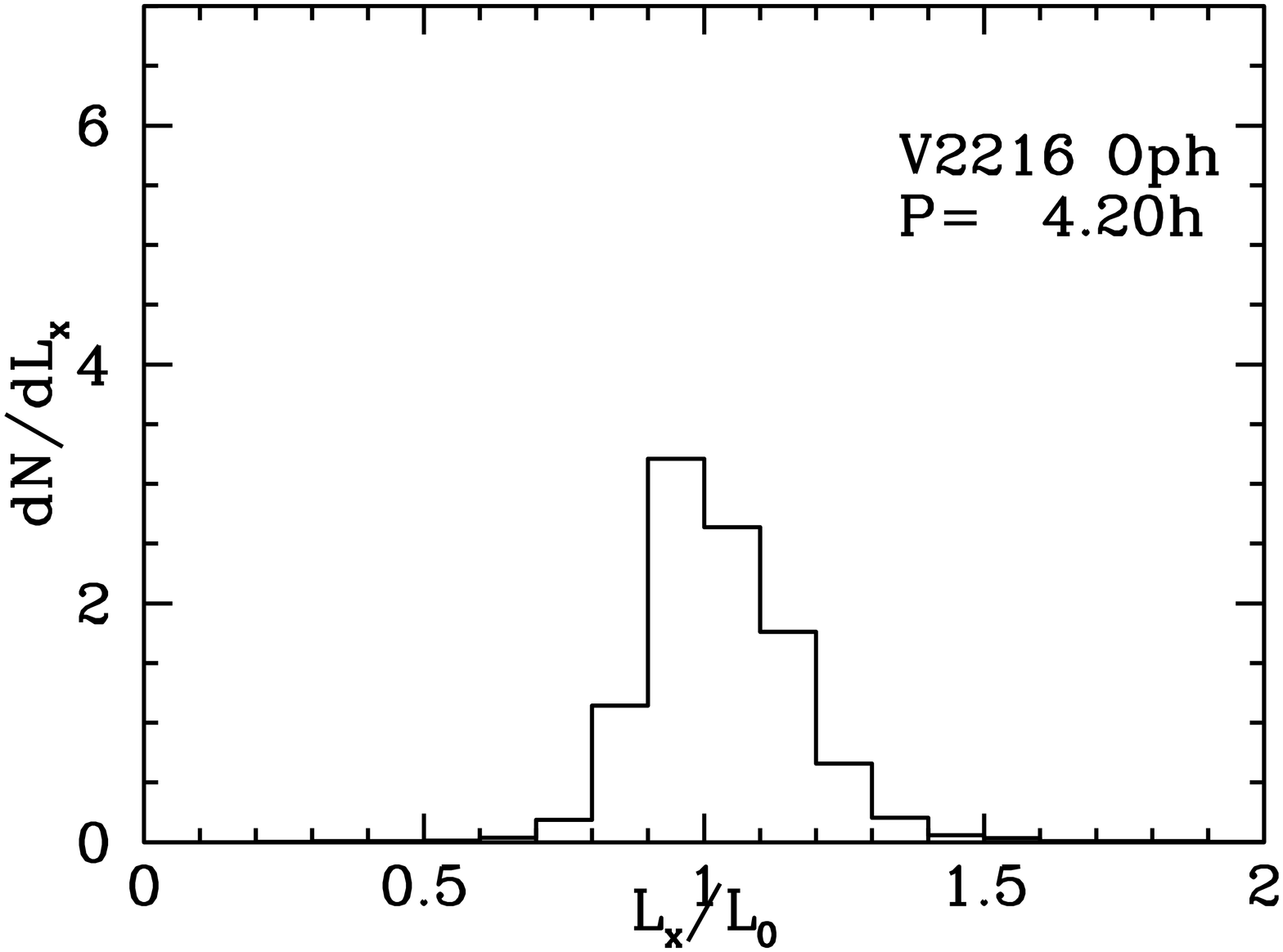}\\
\includegraphics[angle=270, totalheight=1.4in]{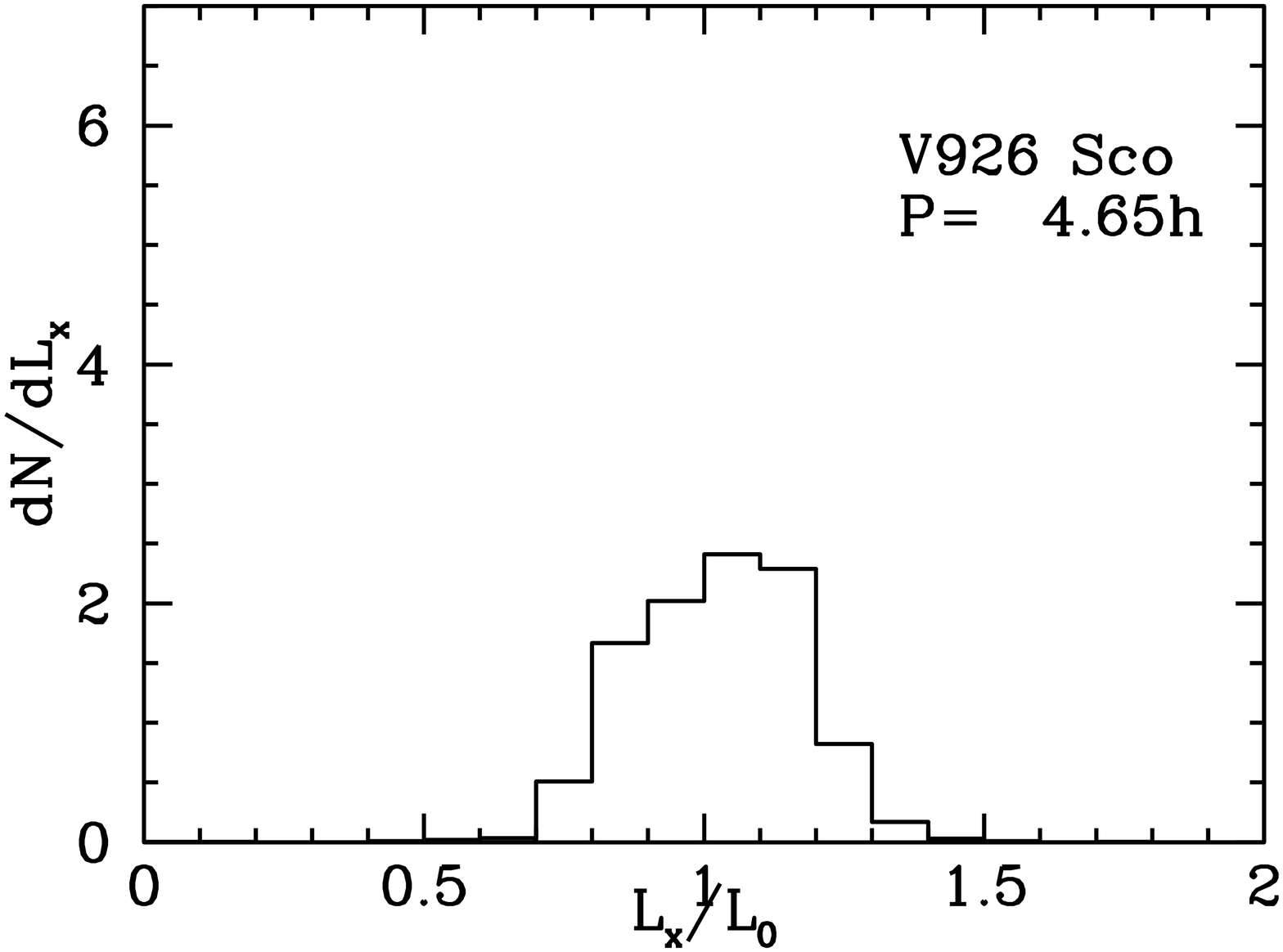}
\includegraphics[angle=270, totalheight=1.4in]{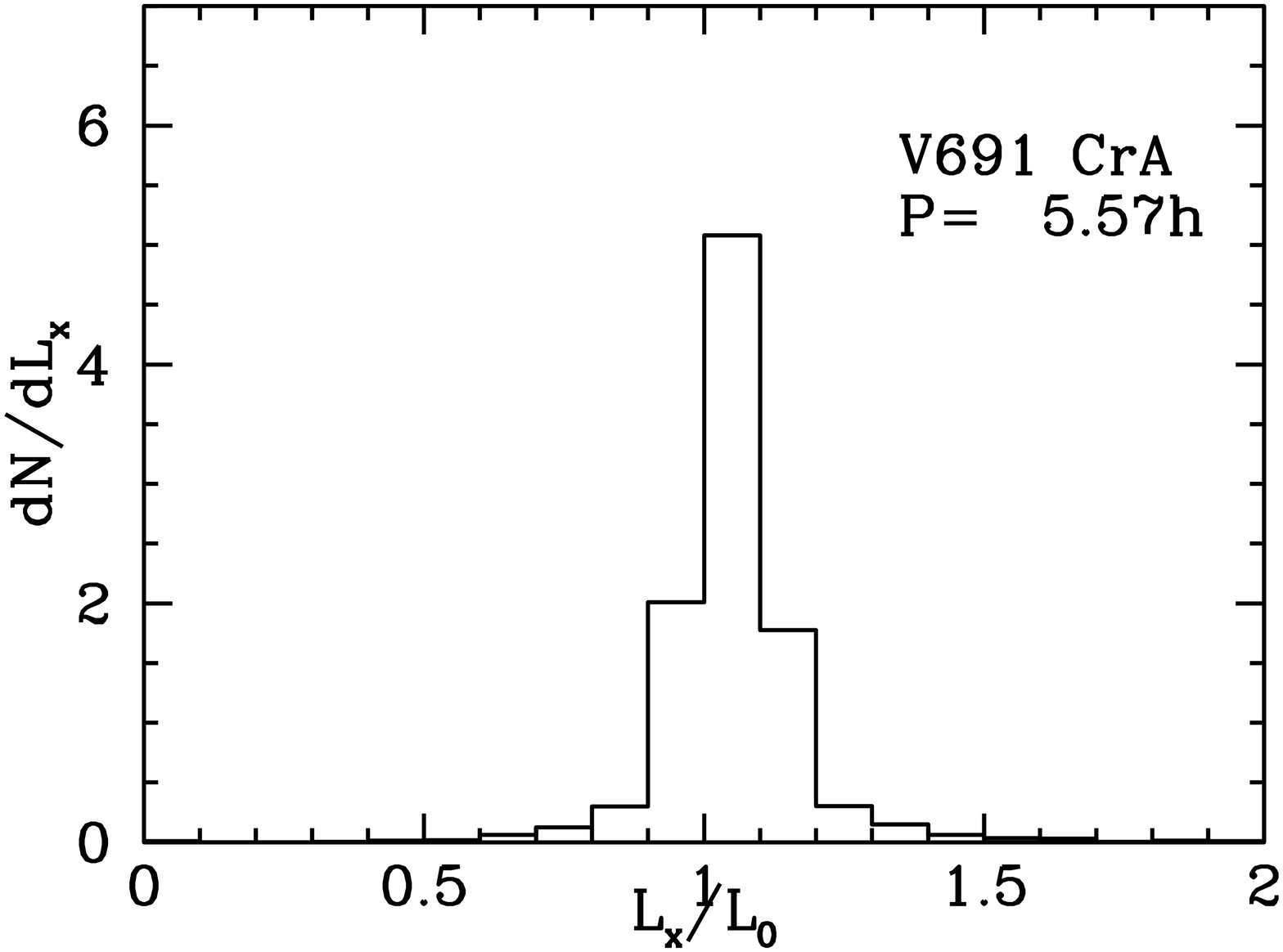}
\includegraphics[angle=270, totalheight=1.4in]{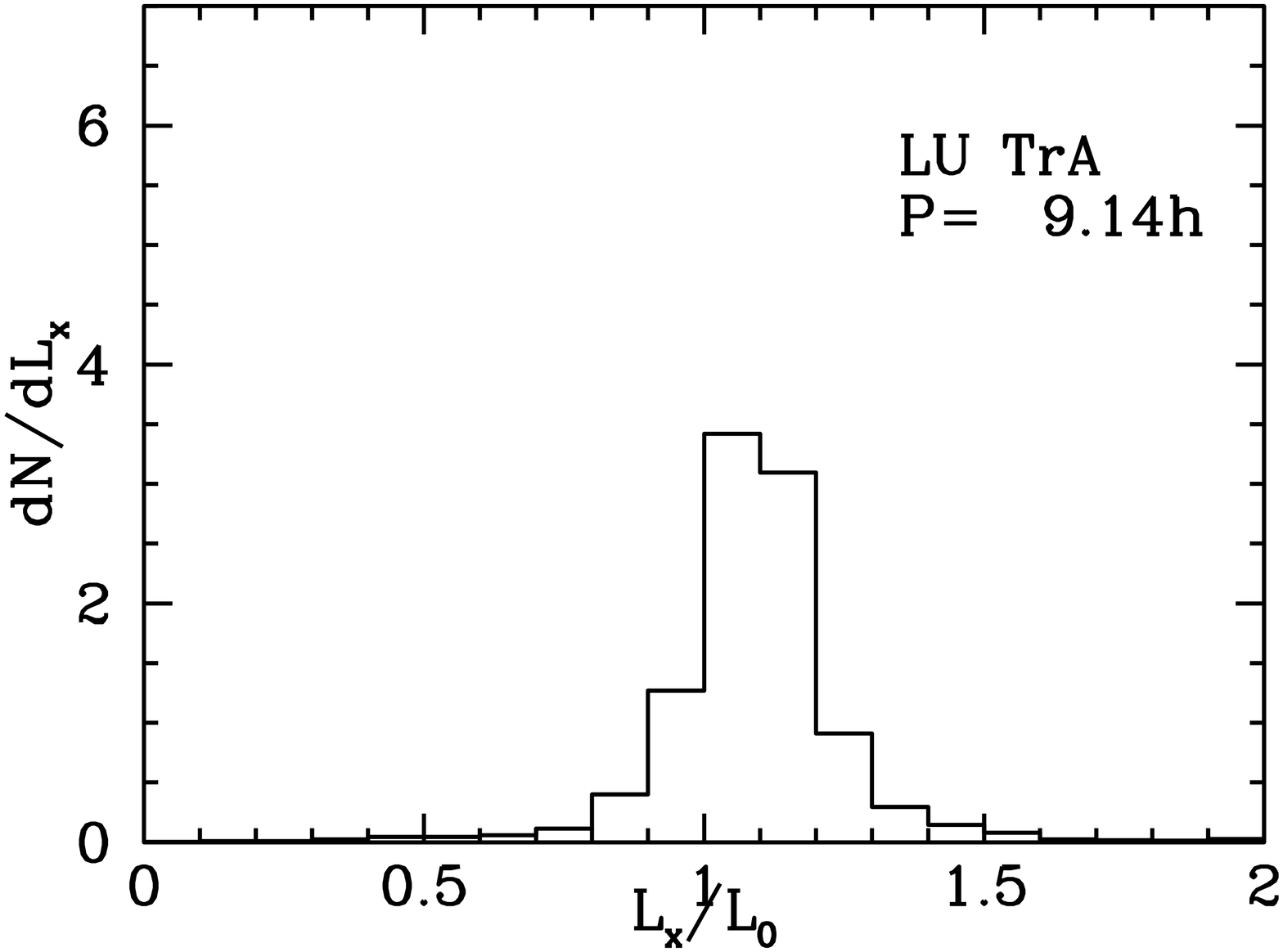}
\includegraphics[angle=270, totalheight=1.4in]{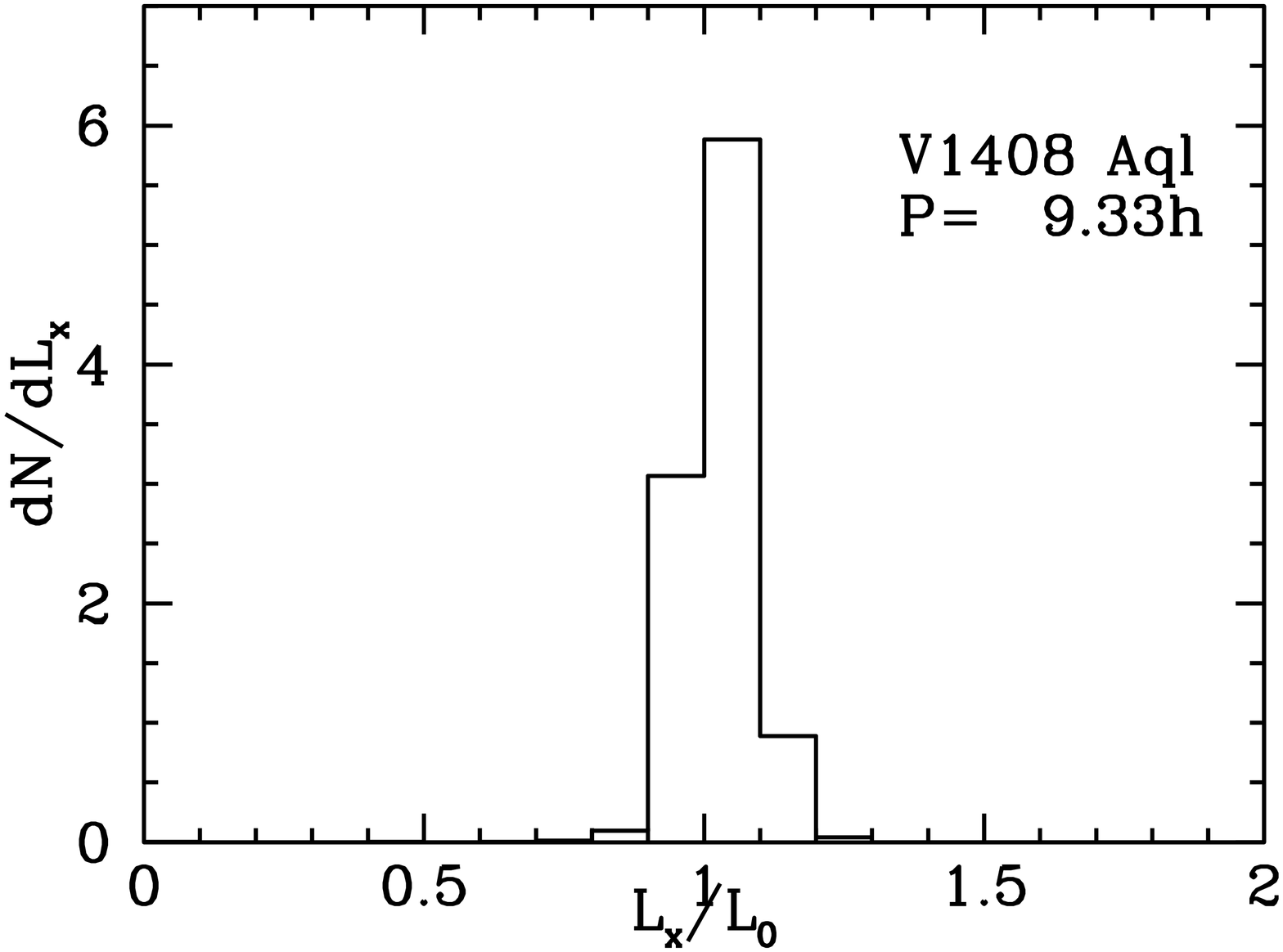}\\
\includegraphics[angle=270, totalheight=1.4in]{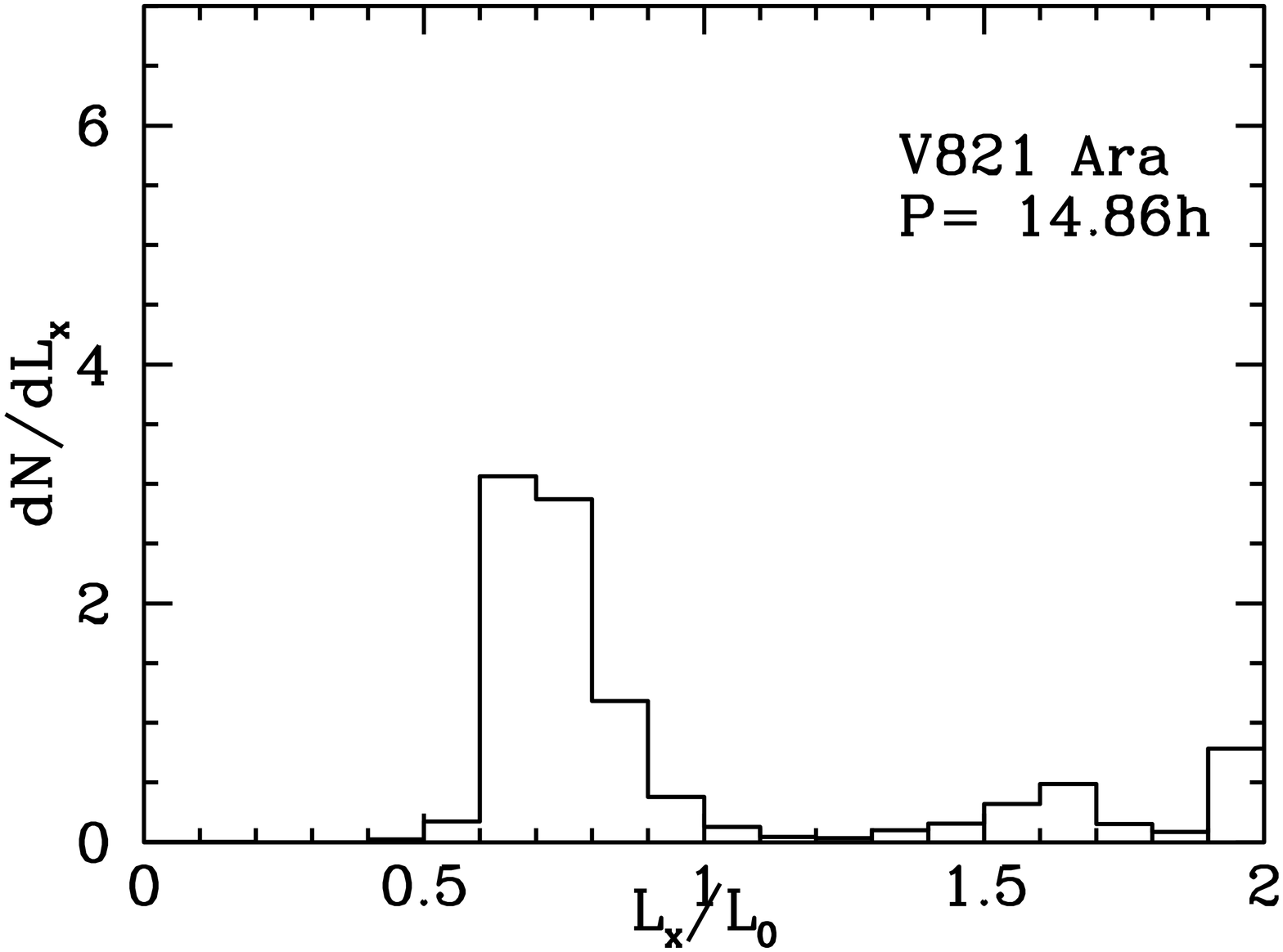}
\includegraphics[angle=270, totalheight=1.4in]{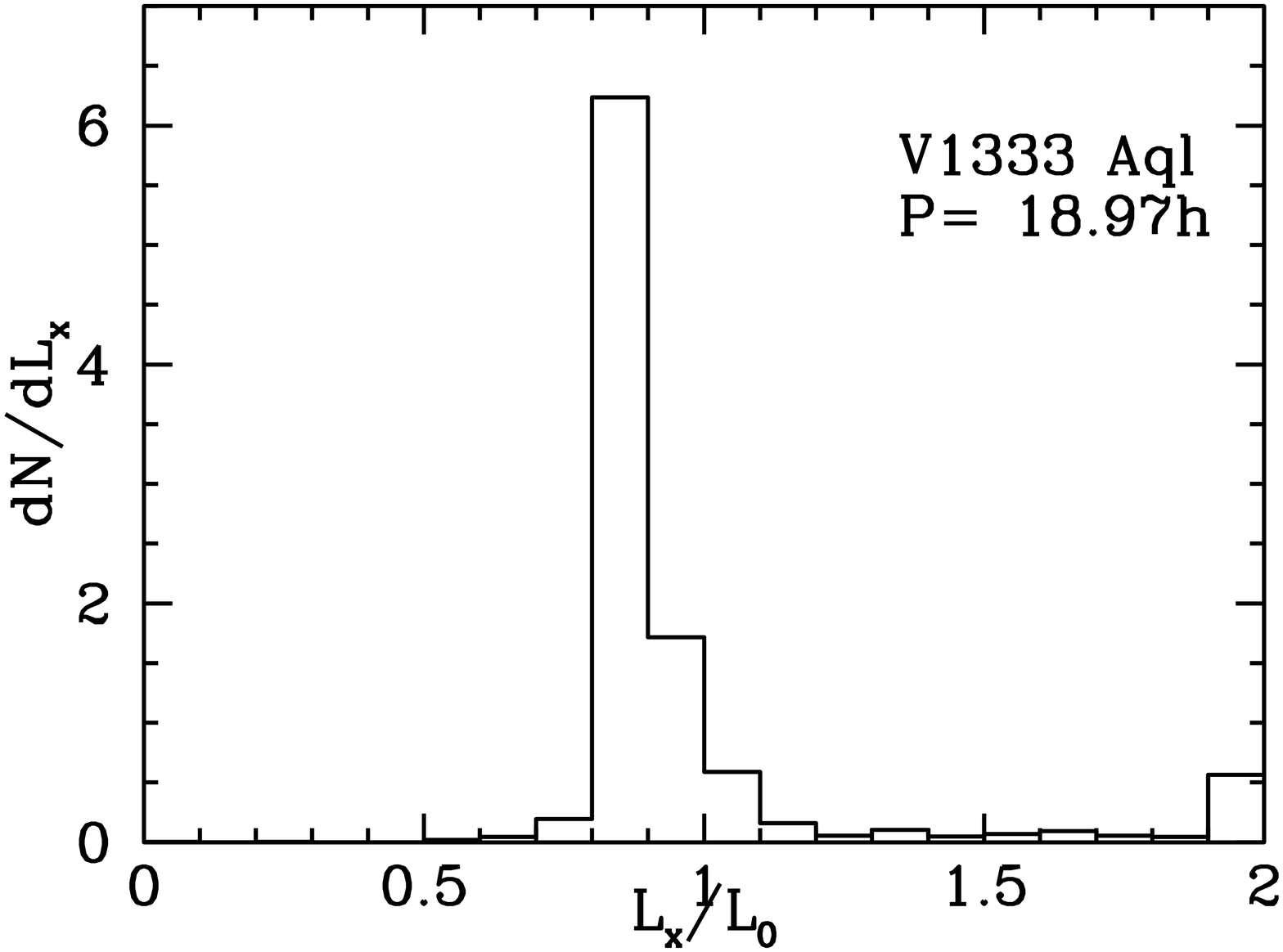}
\caption{XLF for 10 LMXBs from Table 1 (with known orbital
periods shorther than 20 hours) constructed from the 
RXTE ASM data. The luminosity of each source is given in units of 
the mean luminosity 
$L_0$. Distributions are normalized to unity 
\protect{$\int \frac{dN}{dL_x}dL_x=1$}. }                                                   
\end{center}
\end{figure}

Let us use the RXTE ASM data to find the shape of 
the function 
$F(L_x-L_0)$ for galactic LMXBs. First 
let us consider bright LMXBs with known orbital periods
shorther than 20 hours and mean fluxes above 1 ASM count/s
(Table 1). In close binasries with orbital periods below 20 hours
the mass of the optical star filling Roche lobe is 
$\lesssim 2 M_\odot$ (we assume the compact star mass 1.4 $M_\odot$
and main-sequence optical star with solar chemical abundance).
The mass transfer is driven by the orbit angular momentum
removal due to magnetic stellar wind or gravitational wave
emission (for orbital periods shorter than several hours, see below). 
Table 1 lists the mean flux of the source
$F_x$ (averaged over several years of observations, typically
$ 10^3 - 10^4$ individual points per source) and rms deviation 
$\sigma$ from the mean $F_x$. 
Both $F_x$ and $\sigma$ are given in units of the ASM count rate
(we remind that the ASM flux from Crab is 75 ASM counts/s).

In Fig. 1. we plot indivudual normalized XLF of sources from Table 1
calculated with the use of the ASM data and show the values of
the corresponding orbital periods. Fig. 2 shows the mean XLF calculated for 
these systems (the histogram) and its approximation by a quasi-Lorentzian 
curve 
$ F(x-y)=\frac{a}{(x-y)^2+by^2}$ ($a,b=const$), 
which obviously satisfies condition (\ref{F}). 
The errors in each bin are calculated using the formula for
variance
$\sigma_i=\sqrt{\sum_{j=1}^{n_i} (x_j-\bar x)/n_i}$ 
($n_i$ is the number of points int he bin, 
$\bar x$ is the mean value). The formal confidence level 
of the fit according to the $\chi^2$ criterion
for 17 degrees of freedom 
is $ P(\chi ^2_{17}\ge 8.3) \approx 0.96$. 

\begin{figure}
\centerline{
{\epsfig{file=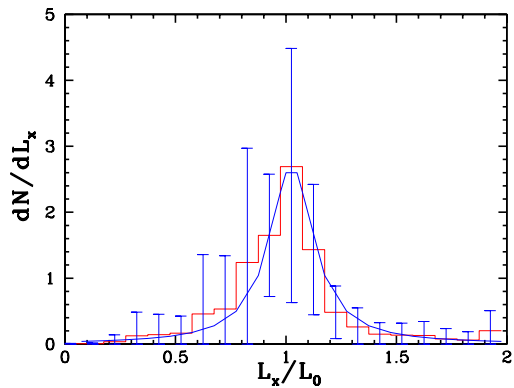,angle=-90,width=7cm}}}
\caption
{The mean XLF for 10 LMXBs from Table 1 (the histogram). The solid lines
shows the approximation by a Lorentzian curve. The confidence level is
$ P(\chi ^2_{17}\ge 8.3) \approx 0.96$.  }
\end{figure}

\begin{table}%[h]
\caption{Bright LMXBs. The mean fluxes 
$F_x$ 
and the rms flux variances  $\sigma$ are given in units
of the ASM count rate. }
\begin{center}
\begin{tabular}{|r|l|c|c|}
\hline
N\% & Source   & $F_x$ & $\sigma$  \\
\hline
\hline
 1 &Cyg X2    & 37.22   &   8.32 \\   
\hline                             
2&Sco X1     &880.995   &114.68\\     
\hline                               
3&GX 9+9     &  17.22  &2.42  \\       
\hline                                  
4&GX 9+1     &  39.24  & 4.57  \\       
\hline
5&GX 5-1     &  70.75  & 9.53  \\        
\hline
6&GX 3+1     &  22.34  & 5.99  \\        
\hline
7&GX 17+2     &  44.86  &5.33  \\         
\hline
8&GX 13+1     &  22.45  & 2.93  \\        
\hline
9&GX 340+0     &  29.39  &4.34  \\   
\hline
10&GX 349+2     &  51.40  &12.19  \\    
\hline
 11 &V926 Sco   &  12.89&   3.38\\
\hline
 12 &V2216 Oph &  17.34&   2.51\\
\hline
 13 &V801 Ara  &  10.90&   4.88\\
\hline
14 &1705-4402  &  11.67&   8.85\\
\hline \hline
\end{tabular}
\end{center}
\end{table}

\begin{figure}
\begin{center}
\includegraphics[angle=270, totalheight=1.3in]{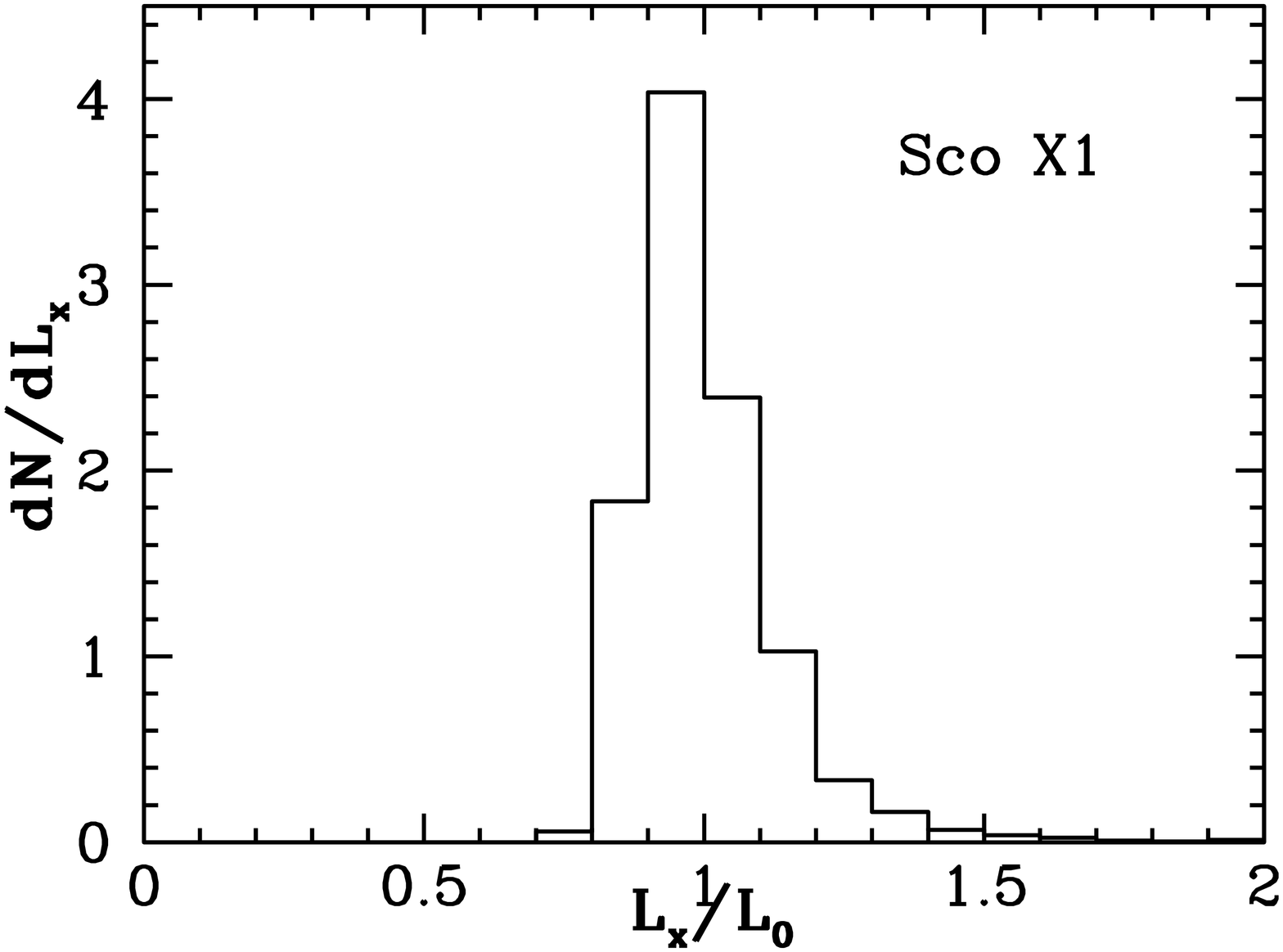}
\includegraphics[angle=270, totalheight=1.3in]{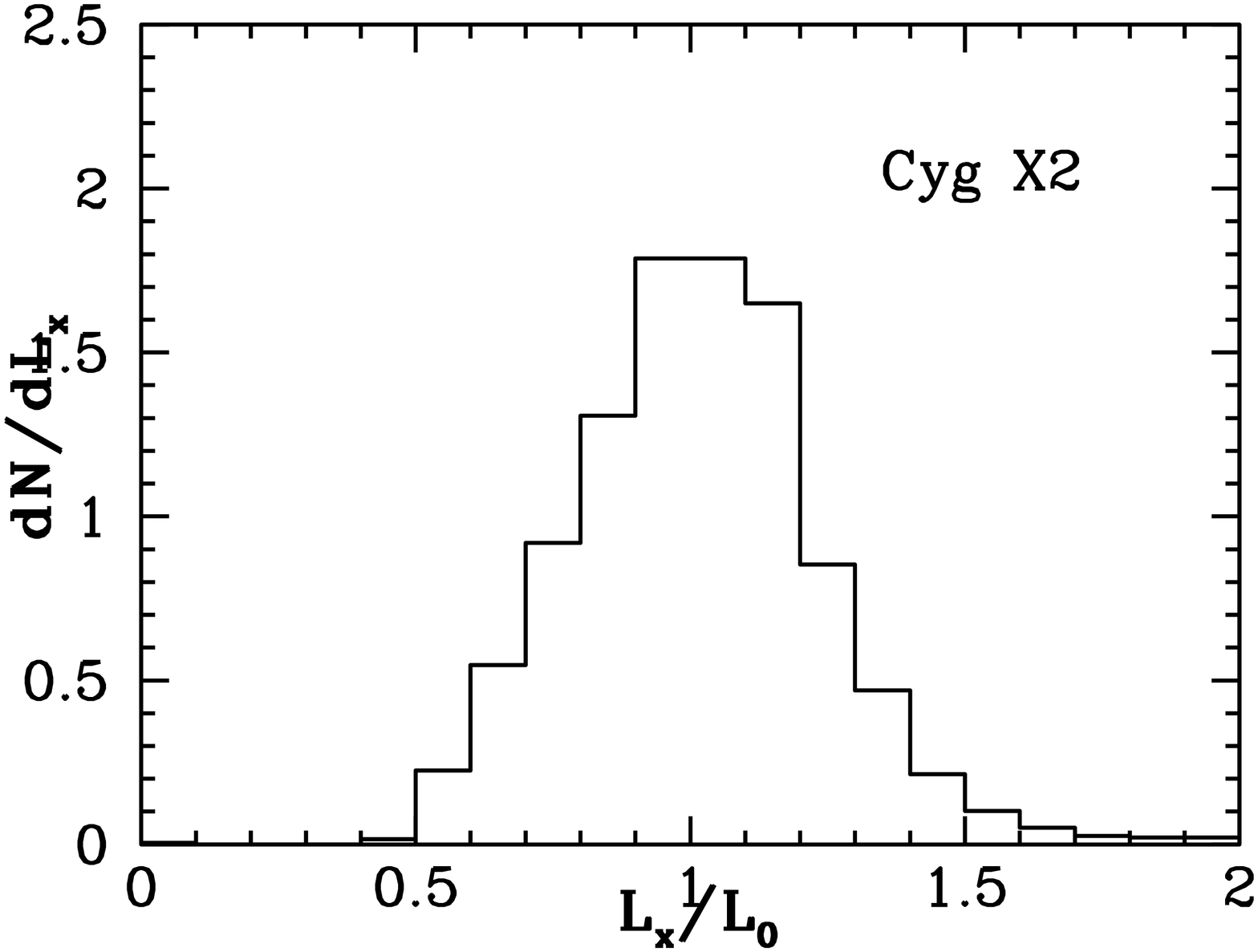}\\
\includegraphics[angle=270, totalheight=1.3in]{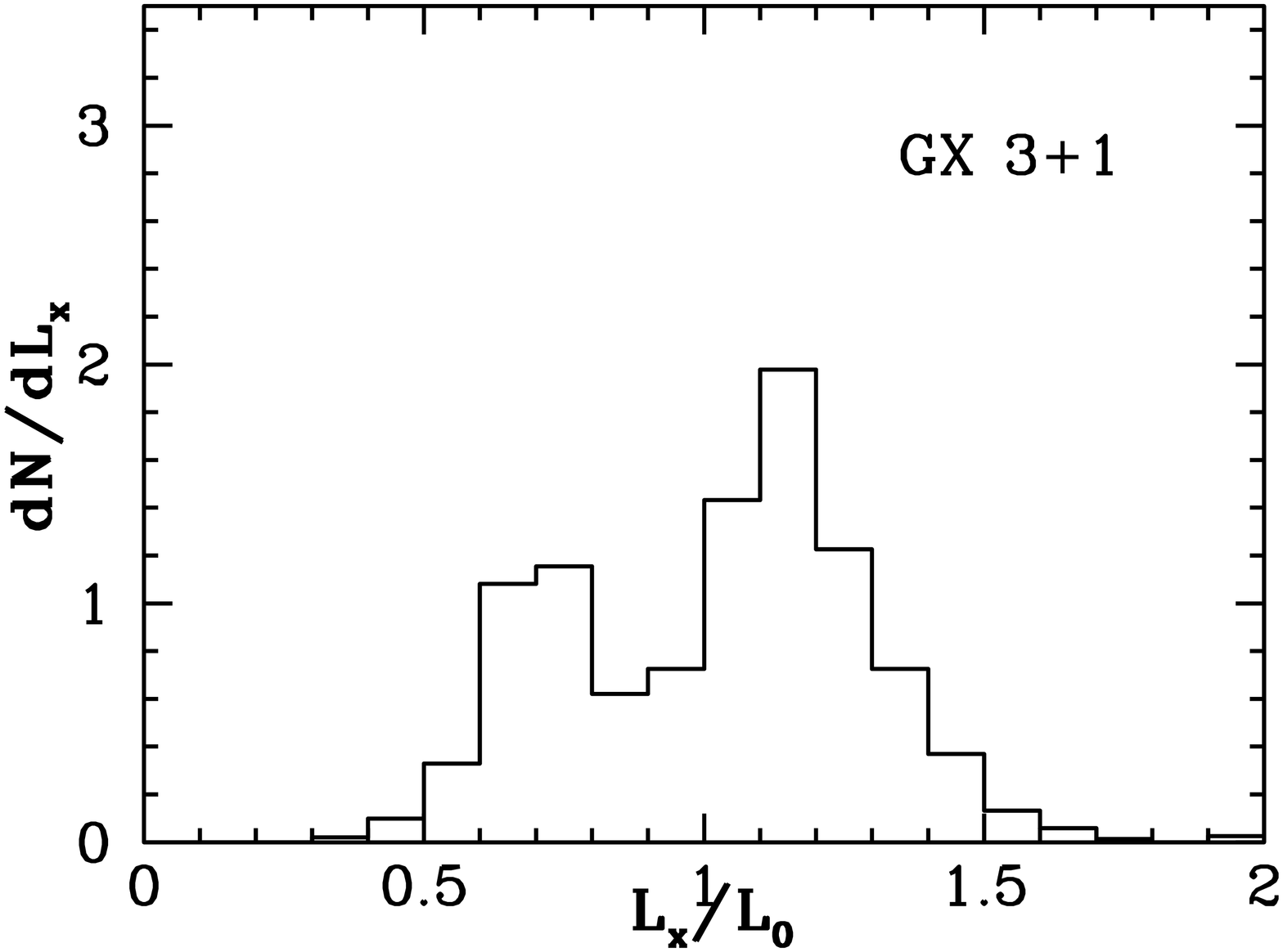}
\includegraphics[angle=270, totalheight=1.3in]{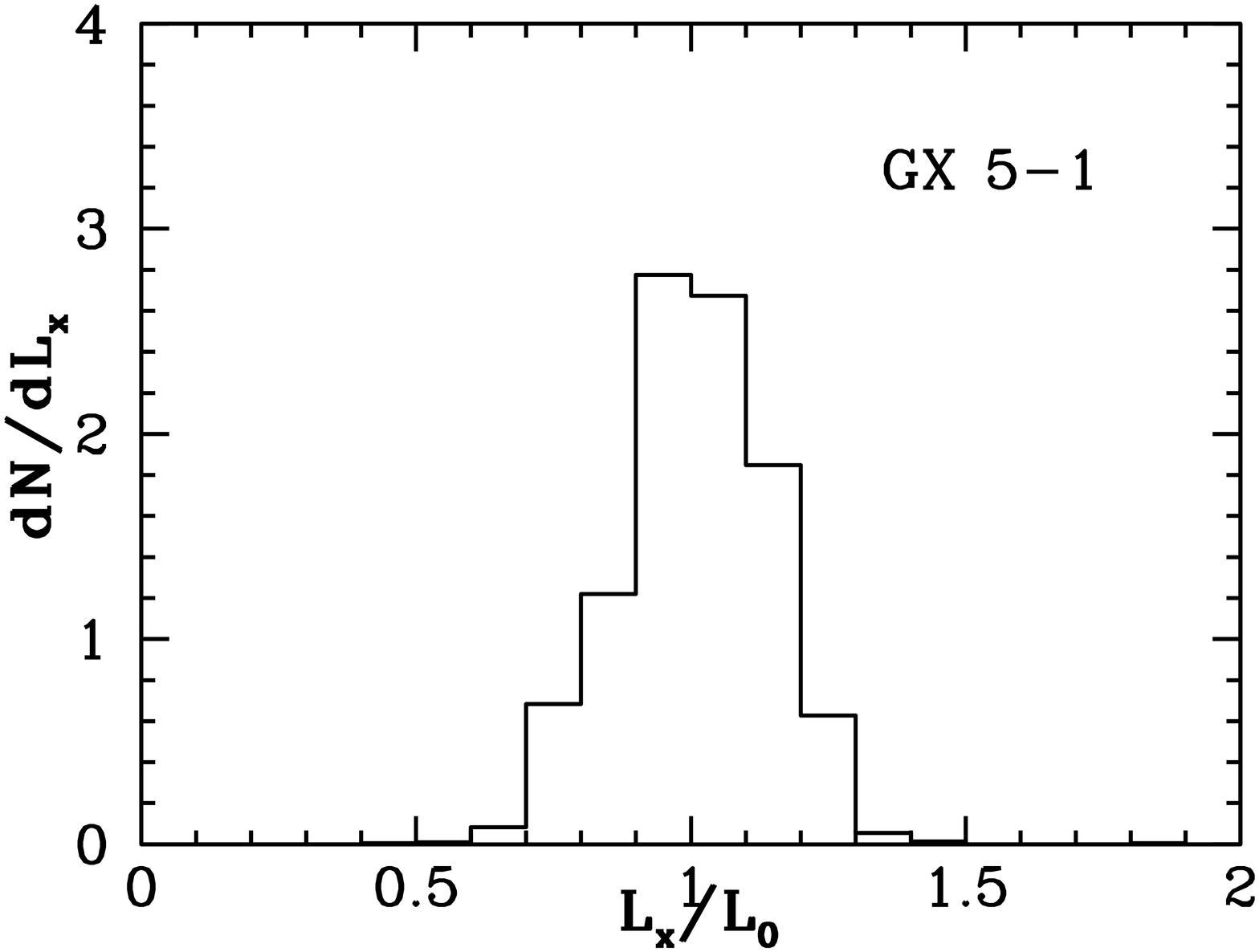}\\
\includegraphics[angle=270, totalheight=1.3in]{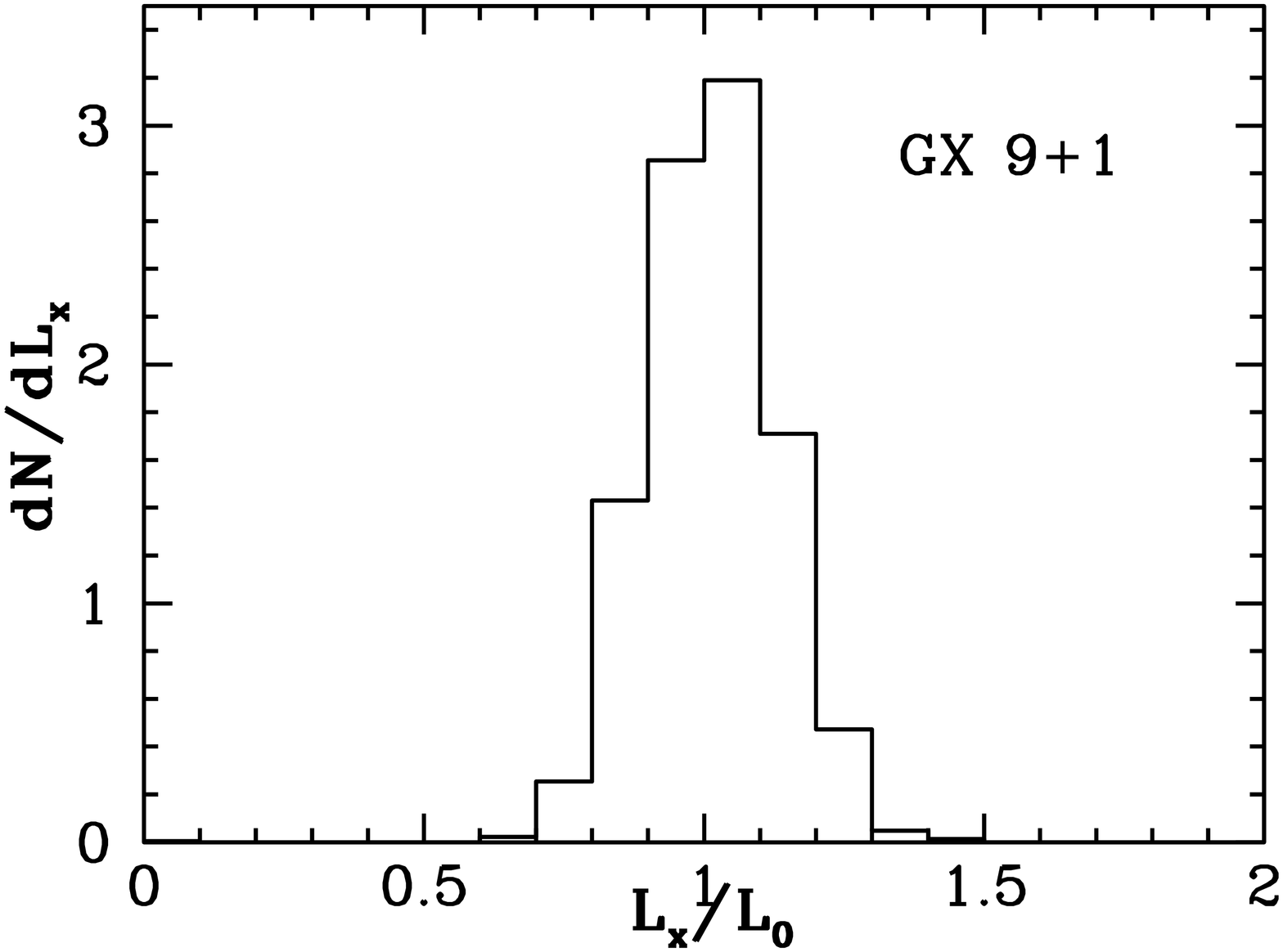}
\includegraphics[angle=270, totalheight=1.3in]{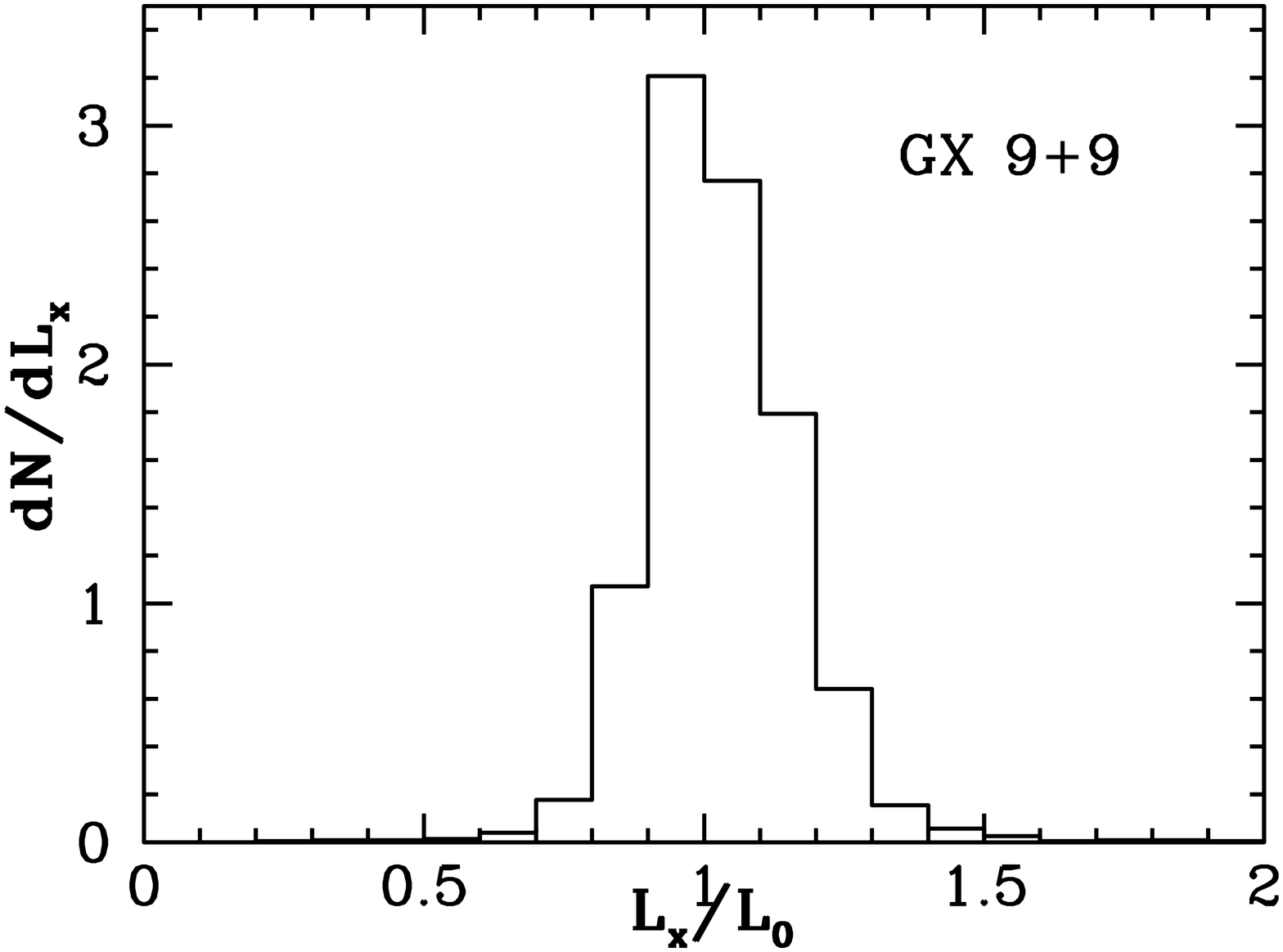} \\
\includegraphics[angle=270, totalheight=1.3in]{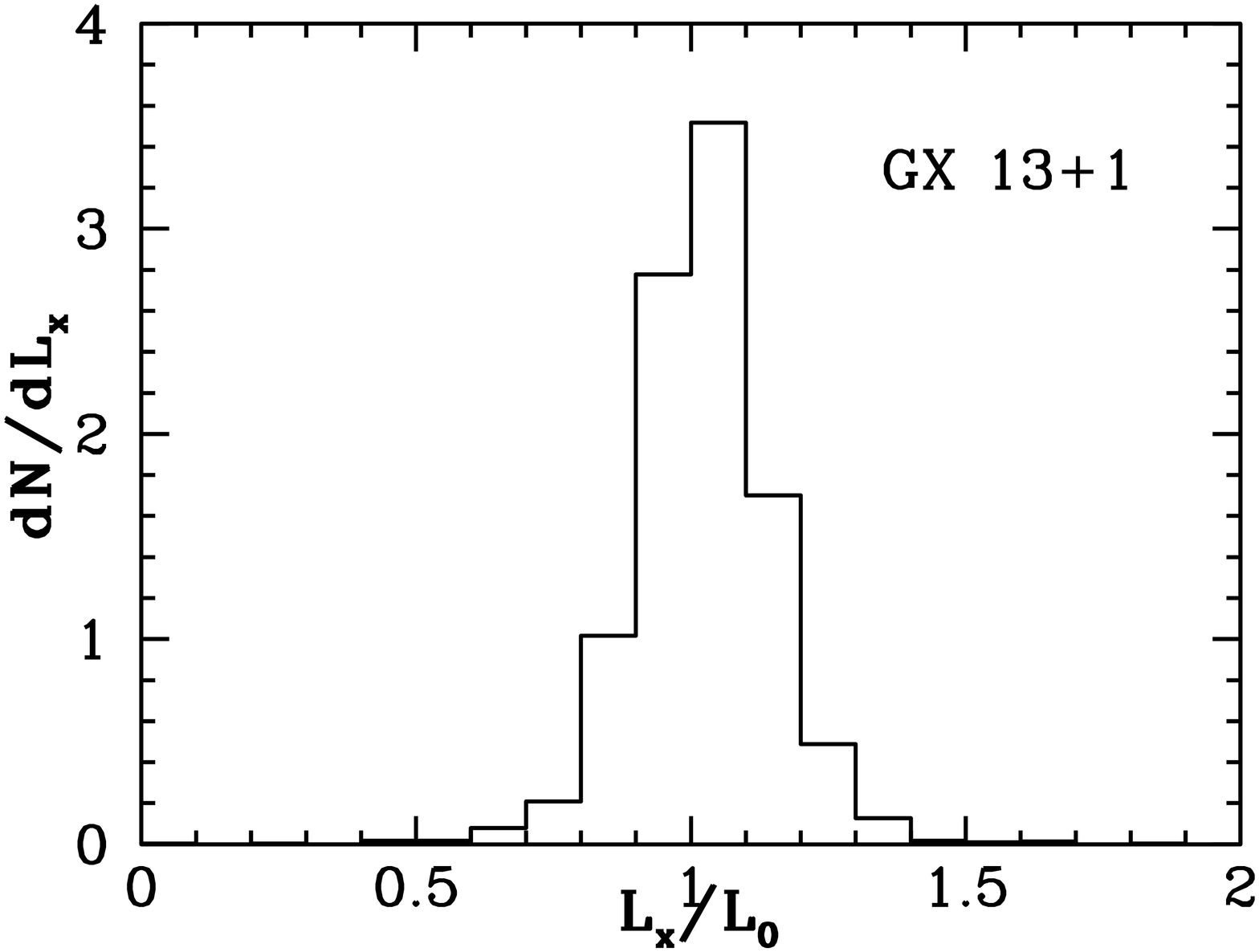}
\includegraphics[angle=270, totalheight=1.3in]{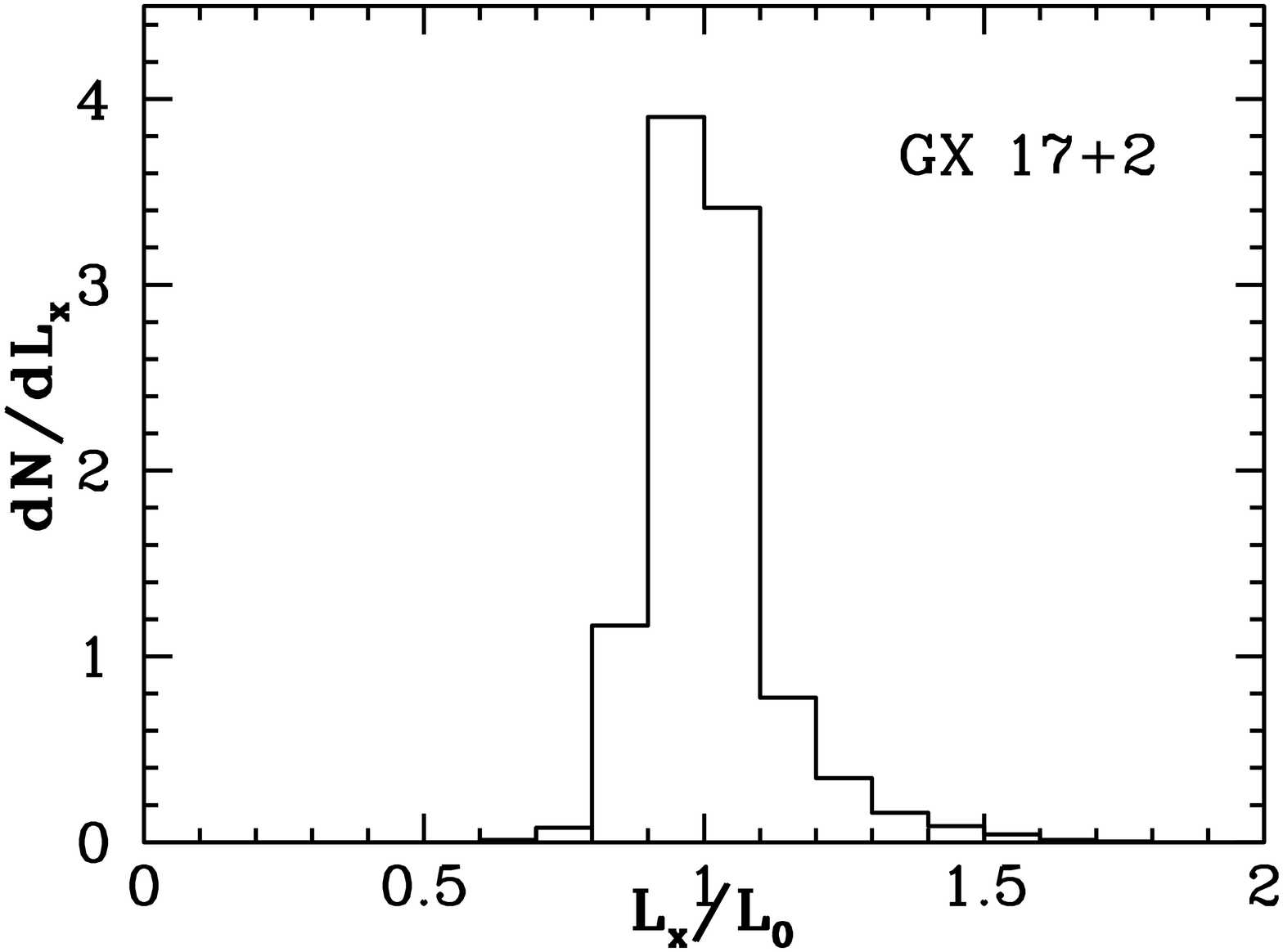}\\
\includegraphics[angle=270, totalheight=1.3in]{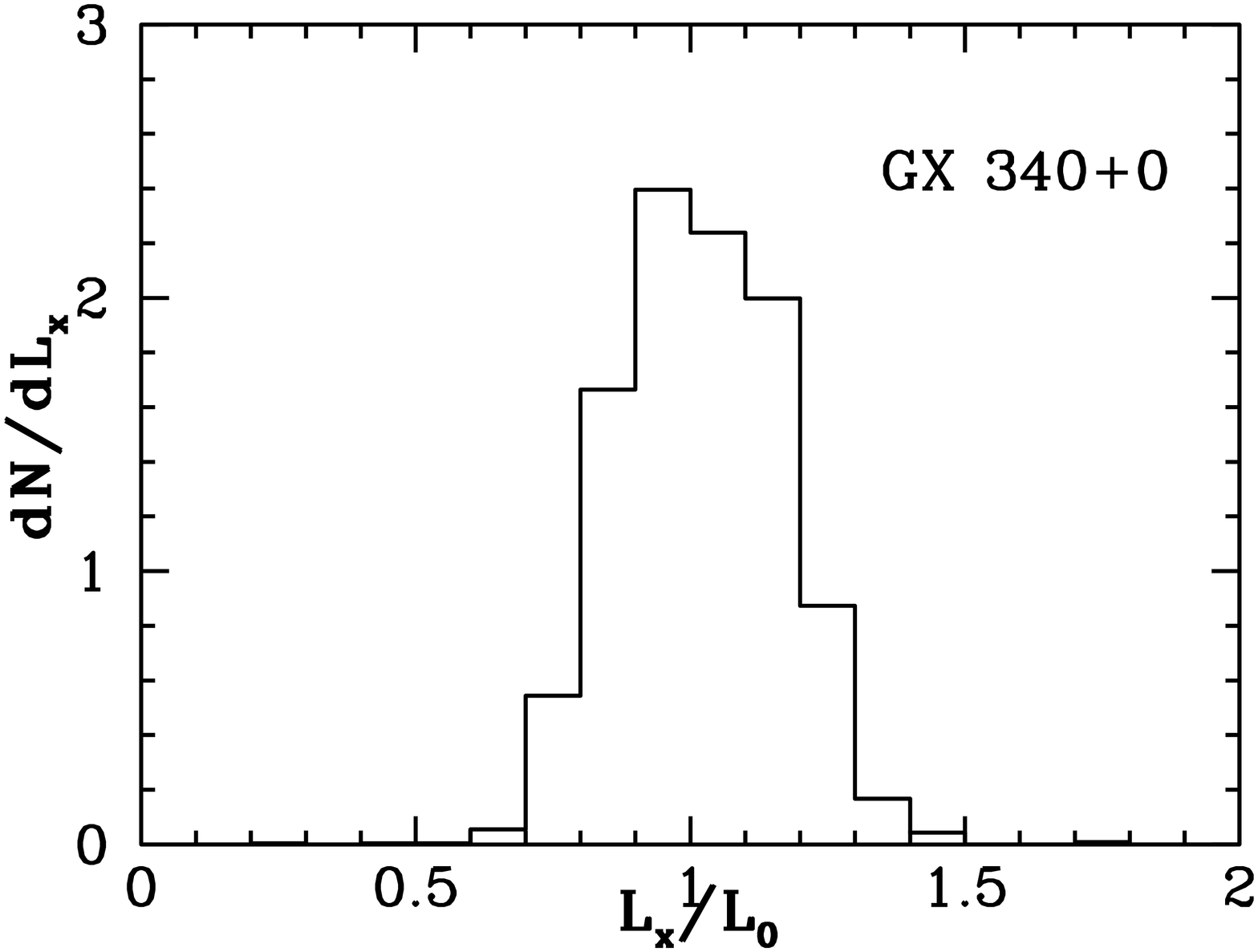}
\includegraphics[angle=270, totalheight=1.3in]{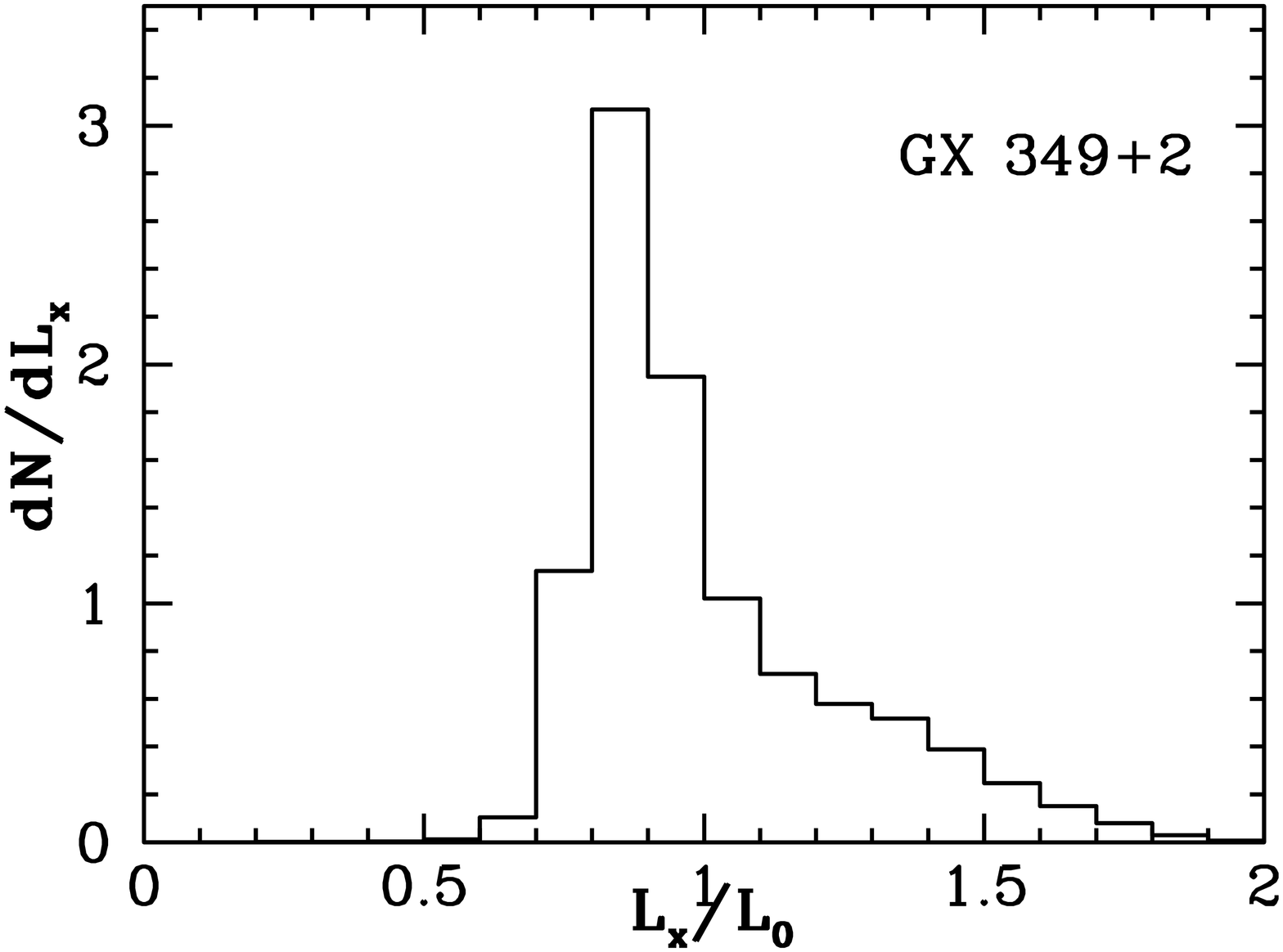}\\
\caption{The same as in Fig. 1 for 10 bright LMXBs from Table 2.  }
\end{center}
\end{figure}

As is seen from Table 1, almost all sources with 
known orbital periods are weak, so individual XLFs (especially 
left wings corresponding to weak fluxes) are determined with 
substantial errors. So as an additional test we used a set 
of bright LMXBs with arbitrary (or unknown) orbital periods (Table 2). 
XLF of an accreting X-ray source is 
determined by properties of non-stationary accretion on the
compact object. The external radius of the accretion disk is 
moderately depends on the system's orbital period
($a\propto P^{2/3}$). The mean flux is determined by the
mass of the compact object. In our case compact objects
are neutron stars of about the same mass. 
Individual XLFs of bright LMXBs from Table 2 
are shown in Fig. 3, the mean XLF for these sources is presented in 
Fig. 4 (the histogram). As for LMXBs with short orbital
periods shown in Fig. 2, the mean XLF for bright LMXBs 
is well approximeted by the Lorentzian curve 
(confidence level $P(\chi^2_{17}\ge 8.8)\approx 0.94$).

\begin{figure}
\centerline{\epsfig{file=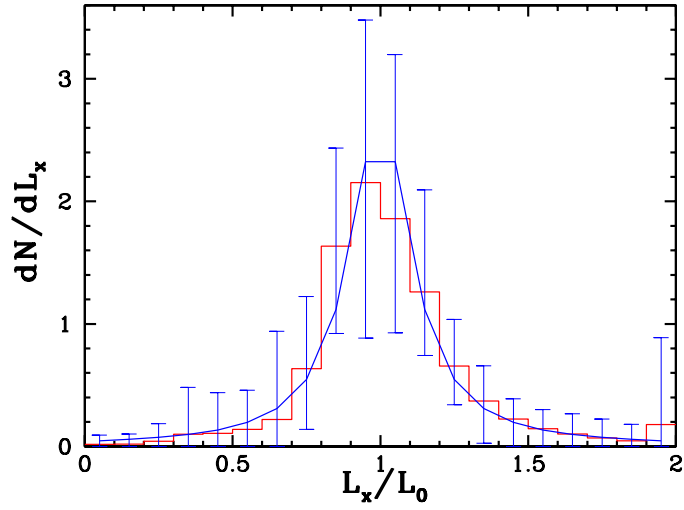, width=7cm, angle=-90}}
\caption{The mean XLF for 14 bright LMXBs from Table 2 (the histogram).
The confidence level is 
$ P(\chi ^2_{17}\ge 8.8) \approx 0.94$.}
\end{figure}

\begin{figure}
\epsfig{file=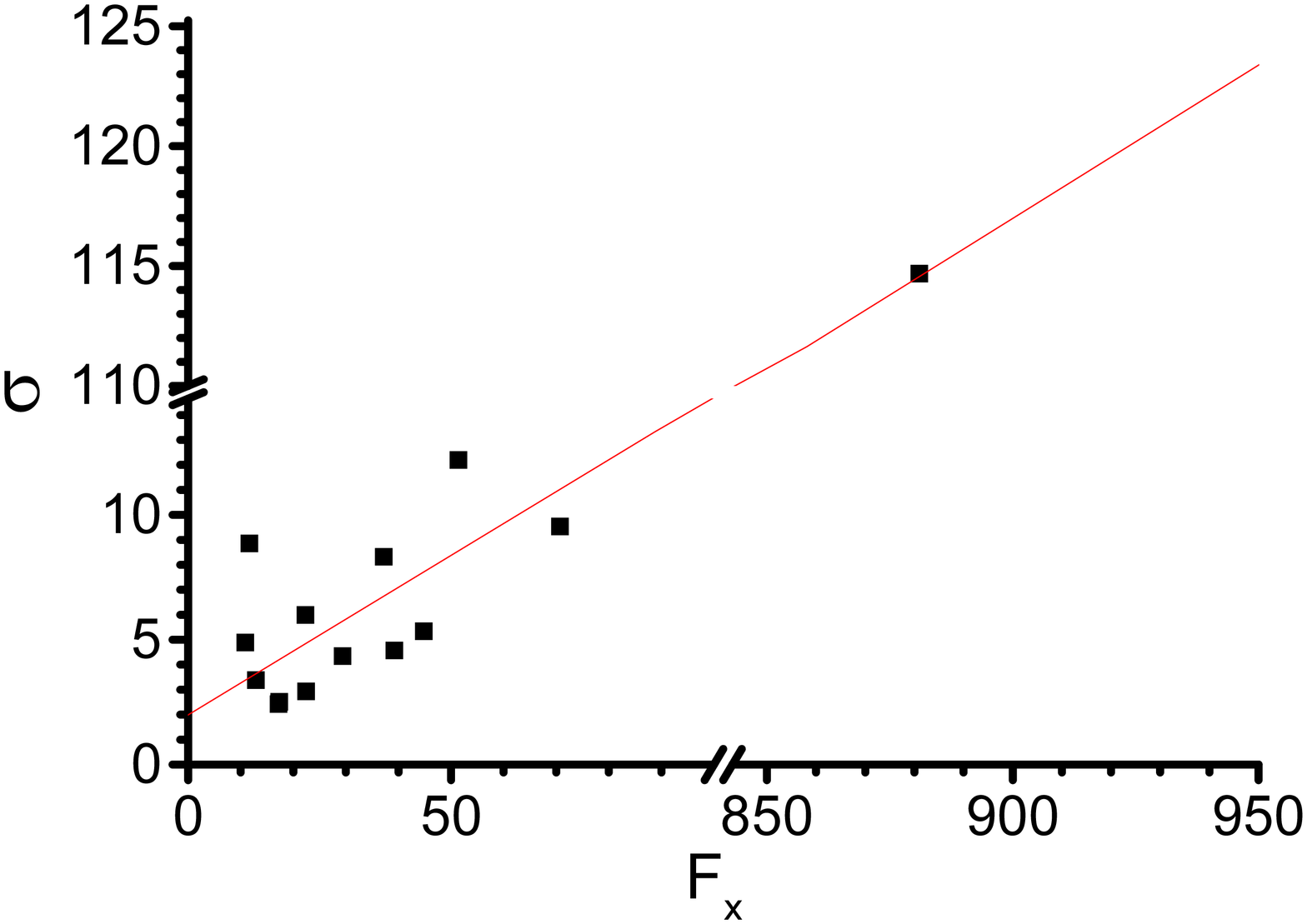, angle=-90, width=7cm}
%        x   x   y    y
%\epsfbox[70 250 550 700]
%{pic5.eps}}}
\caption{The variance of the X-ray flux
$\sigma$ as a function of the mean flux $F_x$ for 14 bright LMXBs
from Table 2. The solid line is the best-fit linear regression. }
\end{figure}

Fig. 5 demonstrates the variance of the observed XLF 
for birght sources from Table 2 as a function of the mean 
flux. The proportionality of the XLF variance to the mean flux 
holds over more than an order of magnitude of flux values.

\section{X-ray luminosity finction of LMXB}

The XLF of LMXB is derived from the  
analysis of observations of point-like sources in bulges
of galaxies (Gilfanov 2004, Kim and Fabbiano 2004) to 
have the power-law from 
$dN/d\ln L_x\sim L_x^{-\Gamma}$
with 
\beq{lf_lmxb_obs}
\Gamma \sim \left\{ \eqalignleft{
\sim 0\,, \qquad & L_x< 2\times 10^{37} \hbox{erg/s} \cr
-0.9\ldots -1.1\,,
      \qquad 2\times 10^{37}< &L_x<5 \times 10^{38} \hbox{erg/s} \cr
-5\,, \qquad &L_x> 5\times 10^{38} \hbox{erg/s}
\cr}
\right.
\eeq
Now we wish to show that for systems with orbital
periods shorther than 
$\sim 20$ hours the power-law form of XLF can be related to principal
evolutionary factors driving the mass exchange in LMXBs -- 
the magnetic stellar wind from optical star 
(MSW) and emission of gravitational
waves (GW). 

\subsection{Magnetic stellar wind}

Let the mass of the optical star and compact object in a binary 
be $M_o$ and $M_x$, respectively, the component mass 
ratio be   $q=M_o/M_x$, and the semi-major axis of 
the binary orbit be $a$. The characteristic time scale of the 
orbital angular momentum loss due to MSW 
$\tau_{MSW}\equiv J/\dot J_{MSW}$
is usually calculated using
the empirical Skumanich law for rotation braking of 
single late-type main-sequence stars, according
to which the surface angular velocity of a star decreases 
gradually with age $t$ as $\sqrt{t}$. 
Applying this law to the late-type low-mass optical component 
in a LMXB and
assuming the tidal locking of the component's spin to the binary
orbital period
(see e.g. Massevich and Tutukov 1988,
van den Heuvel 1992 for more detail) yields
\beq{MSW}
\tau_{MSW}\sim \frac{a^5 M_x}{(M_x+M_o)^2 M_o^4}
\eeq

In LMXBs the mass transfer is due to the Roche lobe
overflow of the optical star so we can write 
\beq{}
R_o=a f(q)
\eeq
where for the shape of the ROche lobe we can adopt 
$$
f(q) \approx \myfrac{q}{1+q}^{1/3}\,, \qquad q<0.5
$$
Then the mean accretion X-ray luminosity is 
$$
L_0\sim \dot M_o\sim \frac{M_o}{\tau_{MSW}}
\sim \frac{M_o^5M_x(1+q)^2f(q)^5}{R_o^5} 
$$
Using the mass-radius relation for late-type main-sequence stars
$R_o\sim M_o^{0.9...1}$ we find
$$
L_0 \sim M_o^{2.17...2.67}, \qquad \alpha\approx 2.17...2.67
$$
Substituting this into Eq. 
(\ref{beta_st}) we obtain 
for the stationary mass function of optical 
comonents in LMXBs
$\beta_{st} \approx 3.52...4$    
assuming the Slapeter intitial
mass function. Finally, we arrive at the luminosity 
function in the form:
\beq{msw_dist}
\frac{dN}{d\ln L_0}\sim L_0^{-1.6...-1.13}\,, \qquad  \Gamma_{MSW}\sim
1.6...1.13
\eeq

\subsection{Gravitational radiation}

In the dipole approximation, the time-scale for the angular
momentum loss in a binary systems  reads
\beq{}
\tau_{GR}\sim \frac{a^4} {(M_x+M_o)M_o M_x}
\eeq
Using the Roche lobe filling condition
$a=R_o/f(q)$ as above we get
$$
L_0\sim \dot M_o\sim \frac{M_o}{\tau_{GR}}
\sim \frac{M_o^2M_x^2(1+q)f(q)^4}{R_o^4}        
$$     
%\v
Utilizing $R_0\sim M_o^{0.9}$ we find 
$$
L_0 \sim M_o^{-0.27...-0.3}, \qquad \alpha\approx -0.27...-0.3
$$
Substituting this into Eq.
(\ref{beta_st}) yields  
$\beta_{st}\approx 1.05...1.08$, so finally we obtain 
the luminosity function 
\beq{gw_dist}
\frac{dN}{d\ln L_0}\sim L_0^{-0.16...}\,, \qquad \Gamma_{GR}\sim
0.16...0.3
\eeq

\subsection{The break of the mean XLF of LMXBs} 

The break of the mean XLF of LMXBs is observed at X-ray luminosities 
$\sim 2\times 10^{37}$ erg/s, which corresponds to the accretion rate on 
the neutron star         
$\dot M\sim 10^{-9}M_\odot/\hbox{год}$.
In our interpretation, the break should correspond to the transition 
from the MSW scale to GW scale as the mass of the optical
star decreases below a certain value where MSW becomes inefficient.
From the simle equality $\tau_{MSW}=\tau_{GR}$ we find the stellar mass 
$$
M_o\approx 0.4 M_\odot
$$ 
which is close to the generally accepted value 
$\sim 0.3 M_\odot$ (Spruit and Ritter, 1983). 
At such masses the mass exchange rate in LMXB drops to 
$$
\dot M(0.4 M_\odot) \sim 3\times 10^{-10} M_\odot/\hbox{yr}
$$
(see, for example, Rappaport et al. 1984 and later calculations
of LMXB evolution with non-degenerate companions), which 
roughly agrees with the observed value of the break. 
 
\section{Discussion and conclusion}

The above analysis allows us to conclude that the observed 
mean XLF of LMXBs can be generally explained by accretion on
neutron stars from Roche-filling non-degenerate component
driven by gravitational wave emission (below $L_x\sim 2\times 10^{37}$ erg/s) 
and by magnetic stellar wind 
(above $L_x> 2\times 10^{37}$ erg/s). 
We have shown that these mechanisms, which determines the evolution
of binaries with orbital periods shorther than 15-20 hours, lead  to 
power-law dependences of 
the mass transfer rate $\dot M_o$ on the optical star mass
$M_o$, 
which is the necessary condition for the power-law form of 
the population of such sources. In LMXBs with longer orbital periods, 
the Roche lobe is filling by a subgiant optical star. 
Using the analytical treatment of the mass transfer process in 
such binaries (Yungelson and Livio 1998),
it is straightforward
to show that in LMXBs with subgiants, too, the power-law dependence 
of the mass exchange rate on the optical star mass is recovered. 

We should also note that a sizable fraction of LMXBs actually recides  
or was born in globular clusters due to effective dynamical
interactions between stars. The evolution of LMXBs in dense stellar
clusters is more complicated than outlined above. So a more accurate
evolutionary treatment of XLF of such LMXBs is required. For example, 
the population synthesis method seems to be promising.

The contribution of ultrashort-period binaries dynamically 
formed in globular clusters (like X 1820-30) into 
XLF of LMXBs was discussed by Bildsten and Deloy (20). In these 
binaries, the Roche lobe is filled by a low-mass degenerate 
white dwarf with inverse mass-radius relation and mass transfer
is controlled by gravitational wave emission. The analysis of these authors
shows that a collection of such ultracompact binaries could significantly 
contribute to the XLF of LMXBs in the luminosity range
$6\times 10^{37} < L_x< 5\times 10^{38}$ erg/s and give the 
slope of XLF as observed. 
                         
Our analysis indicates that the mean XLF of individual 
sources constructed from the RXTE ASM data can be satisfactorily
fitted by a quasi-Lorentzian distribution with variance 
proportional to the mean luminosity in a wide range of X-ray fluxes.
We have shown that such a shape of the mean individual XLF 
does not affect the power-law shape of the XLF formed by 
the entire population of the binary sources. 
The latter is determined by the dependence of the mean 
X-ray luminosity of each source on the mass of the optical star filling 
the Roche lobe.

We stress that it is the possibility to describe the mean XLF
of individual LMXBs by a single law with specific symmetry
(proportionality of the variance to the mean value) that allows
us to recover the power-law form of the XLF of the entire source
population from evolutionary properties of mass transfer in 
LMXBs. Observations show (Kim and Fabbiano 2004) that XLFs of
individual galaxies are somewhat different, the unique poer-law of
XLF is obtained after averaging over many galaxies. This apparently 
reflects individual features of LMXBs in different galaxies
(e.g., various contributions from ultra-compact binaries, globular
clusters, etc.). These observations can be used 
to study formation and evolution of LMXBs in other galaxies. 

\vskip\baselineskip

The authors acknowledge M.Gilafnov and A.Vikhlinin for 
useful discussions and L.R.Yungelson for comments. The 
work is partially supported by RFBR grants 
04-02-16720  and 03-02-16110.
The research of PK was supported by the Academy of Finland
grant 100488.
We made use of RXTE ASM data publically available from HEASARC
of GSFC.

\end{document}